\documentclass{osa-article}

\journal{oe}

\articletype{Research Article}
\usepackage{siunitx}
\begin{document}

\title{High bandwidth laser-frequency-locking for wideband noise suppression}

\author{Mingyong Jing,\authormark{1,2,*} Peng Zhang,\authormark{1,2} Shaoxin Yuan,\authormark{1,2} Linjie  Zhang,\authormark{1,2}, Liantuan Xiao,\authormark{1,2} and Suotang Jia\authormark{1,2}}

\address{\authormark{1}State Key Laboratory of Quantum Optics and Quantum Optics Devices, Institute of Laser Spectroscopy, Shanxi University, Taiyuan, Shanxi 030006, China\\
\authormark{2}Collaborative Innovation Center of Extreme Optics, Shanxi University, Taiyuan, Shanxi 030006, China}

\email{\authormark{*}jmy@sxu.edu.cn} 



\begin{abstract}
Ultra-low frequency noise lasers have been widely used in laser-based experiments. Most narrow-linewidth lasers are implemented by actively suppressing their frequency noise through a frequency noise servo loop (FNSL). The loop bandwidths (LBW) of FNSLs are currently below megahertz, which is gradually tricky to meet application requirements, especially for wideband quantum sensing experiments. This article has experimentally implemented an FNSL with loop-delay-limited 3.5 MHz LBW, which is an order higher than the usual FNSLs. Using this FNSL, we achieved 70 dB laser frequency noise suppression over 100 kHz Fourier frequency range. This technology has broad applications in vast fields where wideband laser frequency noise suppression is inevitable.
\end{abstract}

\section{Introduction}
Laser technology is one of the pillar technologies of modern physics\cite{slusher1999laser}. As modern physics research gradually penetrates more extreme conditions, such as achieving extraordinarily sensitive and accurate measurements\cite{hugi2012mid,gagliardi2010probing} or involving quantum entanglement beyond what is possible classically\cite{aasi2013enhanced,bao2020spin}, it puts forward higher requirements on lasers' performance, especially on its frequency noise characteristics. At present, low-frequency-noise lasers are mainly realized by first taking a stable frequency source, e.g., the mode of an optical cavity processed by ultra-low expansion glass\cite{webster2008thermal} or an atomic transition\cite{pearman2002polarization}, as a reference; second measuring the error between laser frequency and reference frequency by Pound–Drever–Hall (PDH) or other technique\cite{drever1983laser,black2001introduction} and third suppressing the error through a servo circuit\cite{ang2005pid}. Frequency reference, error generation technique, and servo circuit constitute the so-called frequency noise servo loop (FNSL), which is the key ingredient of the active frequency noise cancel system.\par
According to modern control theory, a control system's residul steady-state error inverse proper to system gain, which proper to loop bandwidth (LBW)\cite{ogata2010modern}. Thus, for a closed-loop operation FNSL, the higher the LBW, the lower the laser's residual frequency noise. Lately, the LBW of FNSLs is usually less than megahertz, and there is little improvement due to a lack of demand. The lack of demand mainly comes from the fact that the FNSL with an LBW of hundreds of kHz is sufficient to obtain a mHz linewidth laser based on a low-noise fiber laser\cite{kessler2012sub,matei20171}. This laser satisfies most precision measurement applications that focus only on slowly changing signals (<1 kHz), such as gravitational wave detections\cite{luo2016tianqin,acernese2014advanced} and magnetic field measurements based on the atomic magnetometer\cite{kominis2003subfemtotesla,shah2007subpicotesla}. However, with the rapid development of quantum sensing in recent years, quantum sensors' observation frequency has gradually expanded to higher frequencies\cite{savukov2005tunable,li2020continuous}. One of the typical examples is our recently implemented atomic superheterodyne receiver\cite{jing2020atomic}, where the observation frequency can reach hundreds of kHz. The current FNSL cannot suppress the laser's frequency noise to the shot-noise limit and brings additional noise due to servo resonance within these applications' observation frequency range. Thus, an FNSL with an LBW greater than megahertz is required.\par
It is currently the transducer, which converts a voltage to laser frequency, limiting the LBW of FNSL. When using PZT as a transducer, the LBW is usually below 200 kHz, which is limited by the resonance frequency of PZT\cite{briles2010simple}. The laser's driven current is also a commonly used transducer. Unfortunately, the response of the laser frequency to the drive current suffers from a competition of two phenomena: as the driven current increased, below the reversal frequency (typically several hundreds of kHz), the Joule heating effect is dominant, which tends to lengthen the laser wavelength; on the contrary, the plasma effect is dominant above the reversal frequency, which tends to shorten the laser wavelength\cite{fukuda2010temperature}. This particular response causes a phase reversal at the reversal frequency, making the negative feedback system a positive feedback system, thus limiting the LBW of current-based FNSL less than MHz. AOM is the third commonly used transducer. The loop bandwidth of AOM-based FNSL is usually less than 1 MHz, limited by the time the sound wave needs to cross the beam diameter\cite{484}.\par
This article designed a unique FNSL with loop-delay-limited LBW over megahertz to suppress wideband laser frequency noise. This FNSL combines customization optical and circuit systems and a fiber electro-optic modulator (FEOM) based frequency transducer. The FEOM has a bandwidth over several GHz, limited only by the time the electromagnetic wave needs to cross the fiber core; therefore, it can break through other transducers' limitation on the servo bandwidth. We use this FNSL to lock a noisy diode laser to a reference cavity mode with a loop-delay-limited LBW of 3.5 MHz and achieve a relative linewidth at mHz level. The close-loop residual frequency noise spectral density has 70 dB suppress over 100 kHz.

\section{Experimental setup}
\begin{figure}[h!]
\centering\includegraphics[width=13cm]{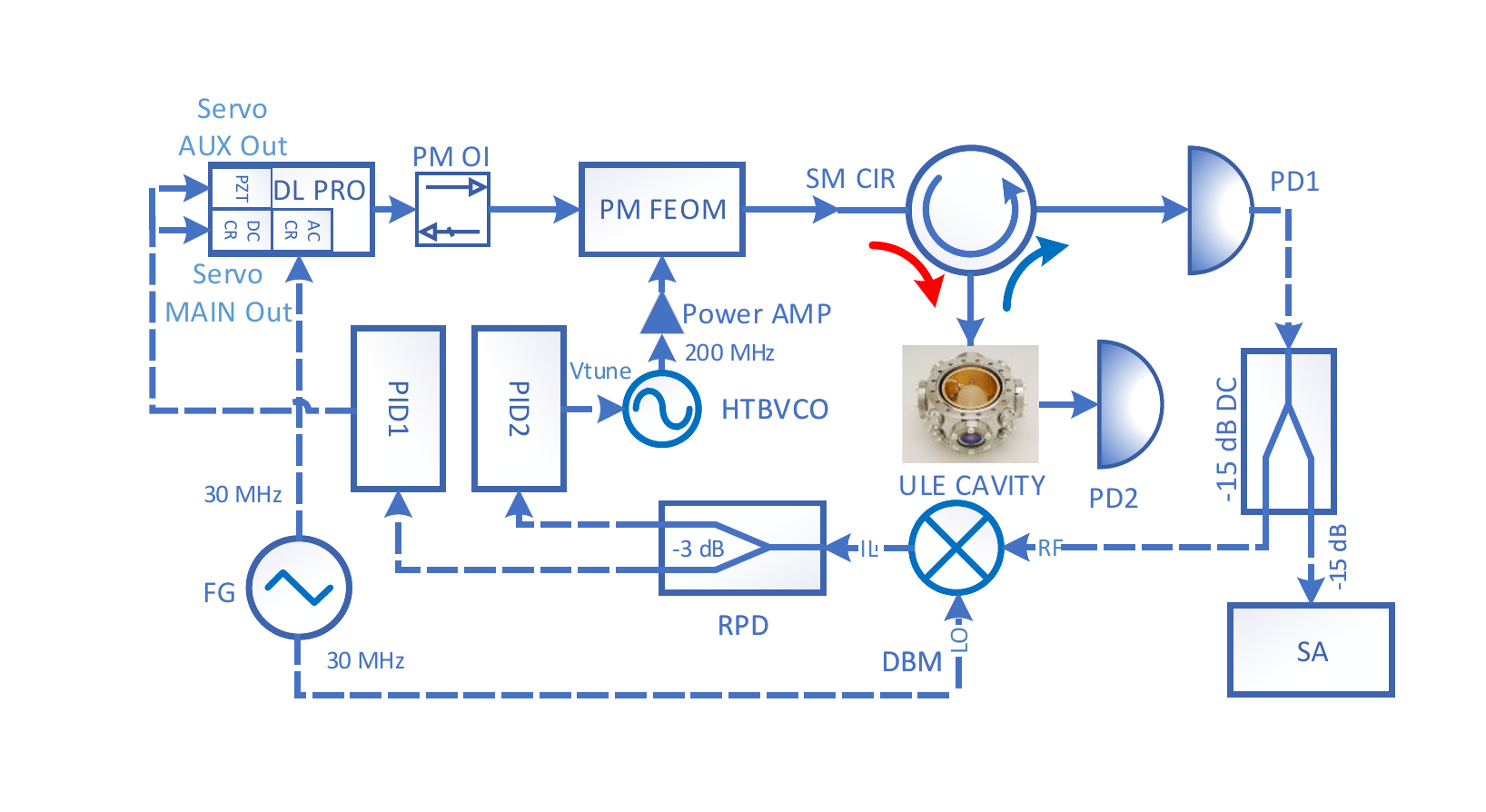}
\caption{\textbf{Overview of the experimental setup.} Solid lines represent optical fibers, and dashed lines represent circuit connections. We have used the following notations: (1) DL PRO: external-cavity diode laser, (2) DC(AC) CR: DC(AC)-coupled laser current port, (3) PZT: piezo ceramic, (4) PM OI: polarization-maintaining fiber optic isolator, (5) PM FEOM: polarization-maintaining fiber electro-optic modulator, (6) SM CIR: single mode fiber optic circulator, (7) ULE CAVITY: ultra-low expansion (ULE) glass cavity, (8) PD1 \& PD2: photodetector, (9) -15 dB DC: -15 dB directional electronics coupler, (10) DBM: double balanced mixer, (11) RPD: 2-way resistive electronics power divider, (12) PID1 \& PID2: proportion integration differentiation controller, (13) HTBVCO: high tuning bandwidth voltage controlled oscillator (200 MHz center frequency), (14) AMP: amplifier, (15) FG: function generator, (16) SA: spectrum analyzer.}
\label{fig:FIG1}
\end{figure}
Figure \ref{fig:FIG1} shows the experimental setup. A commercial external-cavity diode laser emits near-infrared linear polarization light at 1550 nm, which enters the ULE cavity after fiber optic isolator, FEOM and optic circulator. The diode laser have a free-running linewidth about 100 kHz and a polarization extinction ratio about 100:1. Optic isolator and FEOM are processed by polarization-maintaining fiber to minimize laser polarization's influence on the FEOM modulation depth. The ULE cavity is a commercial single-mode fiber fiber-in-fiber-out ultralow-expansion Fabry-Perot (FP) cavity, with a finess of 73,000 at 1550 nm. The linewidth of the high-finesse cavity was about 41 kHz. The cavity was placed in a vacuum system at a residual pressure below $10^{-8}$ mbar, and its temperature was stabilized to the zero crossing point of the coefficient of thermal expansion (\SI{38.7}{\degreeCelsius}). The system was mounted on a passive vibration isolation platform and surrounded by an acoustic and temperature insulating box. A single-mode fiber circulator redirects cavity reflection light to PD1, and the above parts constitute the basic optical path of the PDH technique. The PD1 is an AC-coupled photodetector with a bandwidth over 1 GHz; thus, the detector bandwidth will not limit the LBW. The PD2 is a DC-coupled photodetector to monitor cavity transmission. We use a -15 dB directional electronics coupler to split a small portion (\textasciitilde{}3\%) of the PD1 output signal for measurement with a spectrum analyzer. Most of the PD1 output signal is input to the RF port of the double-balanced mixer and mix with a 30 MHz local signal (LO port) generated by the function generator. The function generator also generates a 30 MHz modulation signal, injecting it into the AC-coupled laser current port to generate PDH optical sidebands. The modulation signal frequency is chosen sufficiently high that Nyquist limit bandwidth is much larger than designed LBW. The modulation signal phase and the local oscillator signal phase can be adjusted separately so that the DBM can demodulate the correct PDH frequency discrimination signal (IF port). A 2-way resistive electronics power divider divides the frequency discrimination signal into two equal parts and injects them into PID1 and PID2 respectively. After PID1 processes the frequency discriminator signal, it outputs the servo voltage through two output ports. One, servo AUX out, has a narrow bandwidth (< 1 kHz) and is used for feedback to the laser PZT, and the other, servo Main out, has a bandwidth of 10 MHz for feedback to the DC-coupled laser current.\par
The second PID, PID2, and FEOM is the key ingredient of high LBW FNSL. The most straightforward configuration is to directly inject the servo voltage from servo Main out of PID2 into the FEOM to achieve a linear phase response relative to the servo voltage, namely linear phase response FEOM based FNSL, abbreviated as LPFNSL. The second configuration that enables FEOM to achieve a linear frequency response relative to the servo voltage is shown in the fig. \ref{fig:FIG1}. In this configuration, we add a voltage controlled oscillator between PID2 and FEOM. The VCO outputs about 200 MHz sine, generates optical sidebands beside the laser carrier. The frequency of optical sidebands relative to laser carrier is equal to VCO driven frequency; thus, is linear response to servo voltage applied on frequency tuning port (Vtune) of VCO. We chose a special high tuning bandwidth (>50 MHz) VCO; thus, the VCO tuning bandwidth will not limit the LBW. The amplifier between HTBVCO and FEOM raises the drive power of HTBVCO, thereby increases the intensity of optical sidebands as much as possible. We name this configuration linear frequency response FEOM based FNSL, abbreviate it as LFFNSL for description convenient. We specify that in LPFNSL configuration, the laser carrier's center frequency is locked to cavity resonance, and in LFFNSL configuration, the upper (or lower) 200 MHz sideband is locked to cavity resonance. In both configurations, cavity transmission can be used as a narrow linewidth laser for subsequent experiments.

\section{Results and discussions}
\begin{figure}[h!]
\centering\includegraphics[width=12cm]{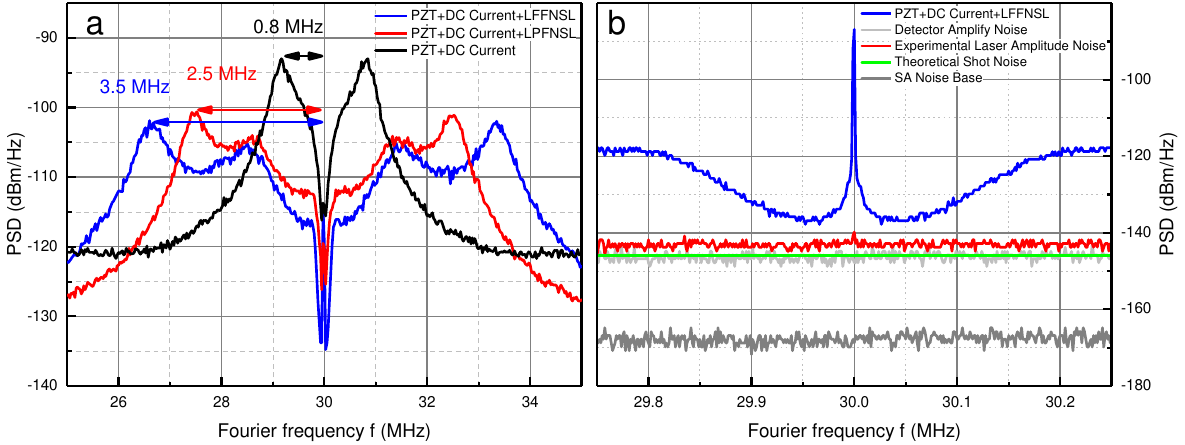}
\caption{\textbf{Comparison of LBW between current-based FNSL, LPFNSL, and LFFNSL configurations.} \textbf{a,} the black, red, and blue curve shows the SA measurement results of close-loop operation current-based FNSL, LPFNSL, and LFFNSL respectively. \textbf{b,} noise analysis of close-loop operation LFFNSL. The gray curve corresponds to the noise base of the spectrum analyzer. The light gray curve corresponds to the amplifier noise of the photodetector. The green line corresponds to the theoretically calculated shot noise level. The red curve corresponds to the laser's experimental intensity noise, and the blue curve is the residue noise of close-loop operation LFFNSL.}
\label{fig:FIG2}
\end{figure}
Figure \ref{fig:FIG2} shows the comparison of LBW of three kinds of configurations: typical current-based FNSL, LPFNSL, and LFFNSL. The LBW is obtained from the closed-loop operation noise spectrum measured by the spectrum analyzer. Around LBW, the FNSL catches a phase reversal and changes from negative feedback to positive feedback. In theory, the overall loop gain of FNSL around LBW should be unit to maintain noise unchanged; thus, the LBW is also referring to unit gain frequency (UGF) of FNSL. However, to obtain a higher low-frequency gain and reduce residual low-frequency noise as much as possible, the loop gain at UGF is often greater than unit in actual experiments, which will cause servo resonance and significantly increase the noise amplitude, appear as a sharper servo resonance peak at UGF in the noise spectrum. As shown in fig. \ref{fig:FIG2}a, after modulation, the servo resonance peaks are symmetrically distributed on both sides of the modulation frequency (30 MHz). The frequency intervals between resonance peaks and modulation frequency read out the LBW. When measures the results of current-based FNSL, only PID1 is working. The current-based FNSL has the narrowest LBW of \textasciitilde{}0.8 MHz, thus the lowest gain and highest residual noise at low-frequency. The LBW of LPFNSL achieves 2.4 MHz, which has about three times improvement than current-based FNSL; thus, LPFNSL suppresses an additional 20 dB Low-frequency noise compare to current-based FNSL. The most powerful configuration is LFFNSL, which has an LBW of 3.5 MHz that 1 MHz wider than LPFNSL, and its residual low-frequency noise is 8 dB lower than LPFNSL. This improvement is mainly because the existing PID circuits are designed for linear response systems and has the best response in LFFNSL. The results also show that the noise rejection ratio, defined as $10\log_{10}(P_{\textrm{open}}/P_{\textrm{close}})$, is roughly proportional to the fourth power of LBW, where $P_{\textrm{open}}$ is the power of free-running noise and $P_{\textrm{close}}$ is the power of close-loop residual noise. For the 3.5 MHz LBW LFFNSL, its low-frequency noise rejection ratio can over 70 dB. Fig. \ref{fig:FIG2}b, shows the noise analysis of close-loop operation LFFNSL. The noise base of the spectrum analyzer is well below the noise of the optic system. The photodetector amplifier noise is at the same level as theoretical shot noise, calculated according to the injected laser power. The experimental laser intensity noise is 3 dB higher than theoretical shot noise due to the incoherence add of detector amplifier noise and optical shot noise. We attribute the additional 10 dB noise gained by close-loop residual noise relative to the shot noise limit to several technique solvable reasons. First, severe residual amplitude modulation (RAM) is caused for we inject the PDH modulation signal to laser current, adding additional noise to the system. Second, without using the balanced detection method, some laser intensity noise is added to the FNSL. Third, the phase noise of FG also contributes to the residual noise.\par
\begin{figure}[h!]
\centering\includegraphics[width=9cm]{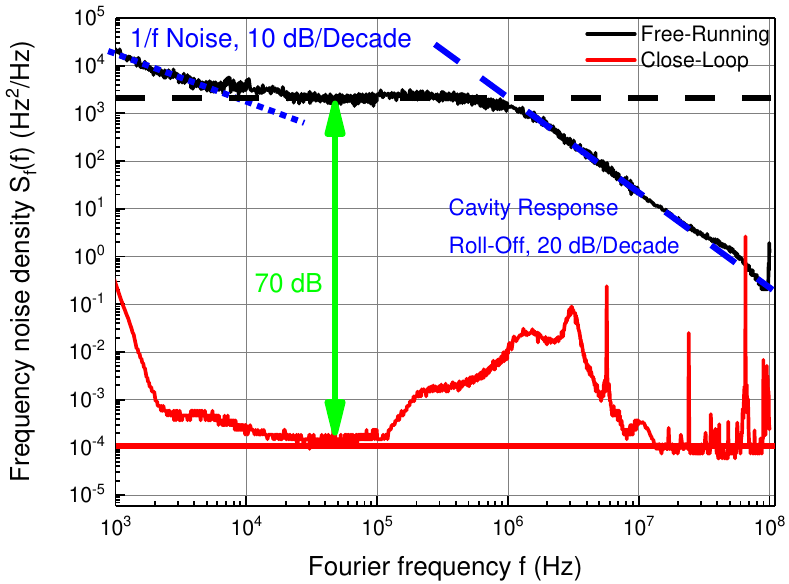}
\caption{\textbf{Frequency noise spectral density (FNSD).} The black curve corresponds to the FNSD of the free-running laser, and the red curve corresponds to the close-loop locked laser. The blue dot line is a fitting of the low-frequency noise of the free-running laser. The blue dash line is a fitting of auxiliary cavity response roll-off. The black dash line and the red line is a fitting of white noise based of free-running laser and close-loop locked laser respectively. Note that a 30 MHz frequency offset is removed and only shows the positive frequency part.}
\label{fig:FIG3}
\end{figure}
Figure \ref{fig:FIG3} shows the frequency noise spectral density (FNSD) of the free-running laser and the close-loop locked laser. The FNSD of free-running laser is measured by the assistance of an auxiliary 1 MHz bandwidth FP cavity. The laser is loose lock to the side of the auxiliary cavity mode with an LBW less than ten hertz to compensate for the longtime drift of the frequency between laser and cavity mode; thus enable longtime stable measurement but hardly influence the noise characteristic of the laser. In this measurement, the auxiliary cavity's transmission is detected by a high-speed photodetector, and the SA processes the output voltage of photodetector. FNSD is calculated by dividing the noise power by a slope, which is obtained by fitting the transmission spectrum of the auxiliary cavity. The FNSD of free-running laser consists of three parts. Below 10 kHz, a 1/f noise with a 10 dB/decade roll-off contributes to the slow linewidth of free-running laser, which is observation time dependent. From 10 kHz to 1 MHz, there is a white noise with a value of $3\times10^3~\textrm{Hz}^2\cdot\textrm{Hz}^{-1}$, which contributes to the 10 kHz fast linewidth of free-running laser. Over 1 MHz, a 20 dB/decade roll-off is shown caused by the auxiliary cavity's integral effect. The FNSD of close-loop locked laser is directly obtained by a calibration process that dividing the residual noise power by the slope of PDH discrimination. The result shows that below 100 kHz, the close-loop locked laser has a white noise base with a value of $2\times10^{-4}~\textrm{Hz}^{2}\cdot\textrm{Hz}^{-1}$, corresponding to a relative linewidth of 6 mHz between laser and reference cavity.\par
\begin{figure}[h!]
\centering\includegraphics[width=12cm]{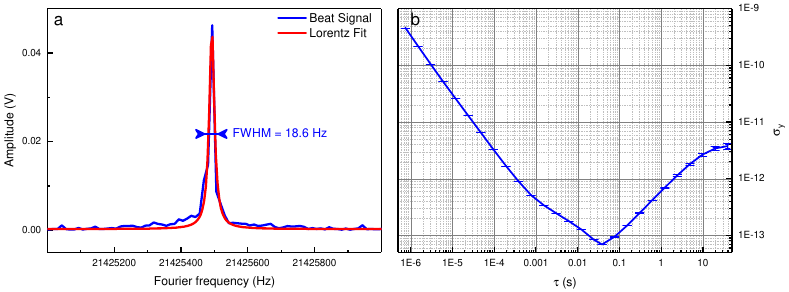}
\caption{\textbf{Beat note of two independent close-loop locked laser systems.} \textbf{a,} Linewidth of beat note signal. The blue curve corresponds to the experimental results, and the red curve corresponds to the Lorentz fit of experimental results. \textbf{b,} Allan variance $\sigma_y$ of beat note signal.}
\label{fig:FIG4}
\end{figure}
Although the 3.5 MHz LBW FNSL can eliminate the relative frequency noise between noisy diode laser and reference cavity to mHz level, the absolute frequency noise of close-loop locked laser systems also depends on the stability of the reference cavity. We beat two sets of  similar laser system to obtain the actual laser linewidth. The two laser systems are independently placed in different laboratories, and the laser are combined through optical fibers, then beat at a high-speed photodetector. The spectrum analyzer reads out the beat signal, as shown in fig. \ref{fig:FIG4}a. Both the resolution bandwidth and video bandwidth of SA are set to 3 Hz, and it takes 5 s to complete a single measurement. The final result is an average of ten measurements with linear frequency drifts in each measurement are removed. The Lorentz fit of experimental results shows the beat note has a linewidth of 18 Hz, which means that a single laser's linewidth is about 13 Hz. Allan variance can also characterize the frequency noise of a laser. We experimentally use a frequency counter to measure the beat note to obtain the Allan variance of the close-loop locked laser, as shown in fig. \ref{fig:FIG4}b. The frequency counter was synchronized with a 10 MHz GPS disciplined oscillator with rubidium timebase (GPSDO) to minimize its long-time frequency drift. The result shows that the close-loop locked laser system has a minimal Allan variance of $7\times10^{-14}$ at an integration time of 38 ms. We attribute this limitation of laser linewidth and Allan variance to the cavity length change caused by thermal fluctuations or mechanical vibration. \par
\begin{figure}[h!]
\centering\includegraphics[width=9cm]{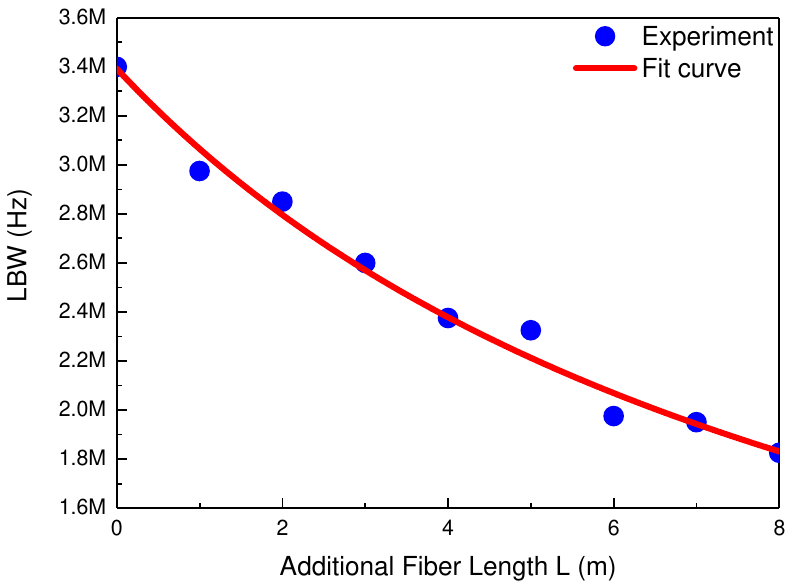}
\caption{\textbf{LBW versus loop delay.} The blue dot corresponds to the experimental results, and the red curve corresponds to the fit of experimental results.}
\label{fig:FIG5}
\end{figure}
We experimentally studied the influence of loop delay on LBW. In this experiment, we add different lengths of additional fiber into the servo loop and measure the LBW respectively. The experimental results are shown in fig. \ref{fig:FIG5}. We fit the experimental results by the equation
\begin{equation}
    \text{LBW}=\frac{C}{2\pi n (L_0+L)},
\end{equation}
where $C$ is the light speed, $n=1.468$ is the refractive index of 1550 nm optical fiber, $L$ is the independent variable, which means the length of the additional fiber. $L_0$ are used as a fit parameter and means the original loop length of our LFFNSL, including the lengths of optical fiber devices, coaxial cables, and equivalent lengths of circuit devices. The fit result shows $L_0=9.38~\text{m}$, is consistent with our experimental settings. The results show that the loop delay ultimately limits the current LBW of our LFFNSL, and the use of shorter pigtail fiber device or opitcal devices on chip to reduce the optical path can further increase the LBW. The second main factor limiting LBW is the 10 MHz bandwidth of the PID circuit, which can be solved by designing a high-speed circuit. Theoretically, in our FEOM-based LFFNSL, the transducer's bandwidth can be above several GHz, which will almost no longer limit the LBW.
\section{Conclusion}
We experimentally realized high-bandwidth LFFNSL. The LFFNSL configuration has achieved a 3.5 MHz LBW limited by the loop delay, which is about an order of magnitude higher than the conventional FNSLs. At present, conventional FNSLs can implement mHz-level lasers based on low-noise fiber laser systems that only require small servo bandwidth, but noisy diode lasers are beyond their touch. Using our LFFNSL, we have achieved a 70 dB frequency noise rejection ratio over several hundreds of kHz, which is sufficient for locking a noisy diode laser to mHz level; therefore, it can be used to realize a narrow linewidth diode laser system at mHz level. This bandwidth can further enhance by simple technique improvements to meet the higher requirement. Although we realized this LFFNSL at 1550 nm laser system, this method can be realized at any wavelength at which the FEOM is mature; thus, this method can have a wide application in atomic quantum sensor experiments or any laser frequency sensitive precision measurement experiments.

\section*{Funding}

\section*{Acknowledgments}
We acknowledge the National Natural Science Foundation of China (grants 61827824, 61475090) and the Science and Technology on Electronic Information Control Laboratory. L.X. acknowledges support from the Program for Changjiang Scholars and Innovative Research Team (grant no. IRT13076).

\section*{Disclosures}

The authors declare no conflicts of interest.


\bibliography{sample}






\end{document}